\documentclass[aps,prx,twocolumn,superscriptaddress,floatfix]{revtex4-1}
\usepackage{amsmath}
\usepackage{amsfonts} 
\usepackage{graphicx}
\usepackage{verbatim}
\usepackage{natbib}
\usepackage{threeparttable}
\setcitestyle{super}

\begin{document}
\title{Longitudinal polaritons in crystals}
\author{Eduardo B. Barros}

\affiliation{Department of Physics, Federal University of Cear\'a, Fortaleza, Cear\'a, 60455-760 Brazil}

\author{Stephanie Reich}
\affiliation{Department of Physics, Freie Universität Berlin, 14195 Berlin, Germany}

\date{\today}

\begin{abstract}
The collective excitations of solids are classified as longitudinal and transverse depending on their relative polarization and propagation direction. This seemingly formal classification results in surprisingly distinct types of excitations if calculated within the Coulomb gauge. Transverse modes couple to free-space photons and hybridize into polaritons for strong light-matter coupling. Longitudinal modes, in contrast, are seen as pure matter excitations that produce a dynamic polarization inside the material without photon coupling. Here we show that both longitudinal and transverse modes become polaritons in the explicitly covariant Lorenz gauge. Longitudinal excitations couple to longitudinal and scalar photons, which have been considered elusive so far. We show that the dipolar excitations become three-fold degenerate in the long-wavelength limit when including all photonic degrees of freedom, as expected from symmetry. Our findings demonstrate how choosing a gauge determines our thinking about materials excitations and how gauge fixing reveals new pathways for tailoring polaritons in crystals, metamaterials, and surfaces. Longitudinal polaritons will interact with longitudinal near fields located at surfaces, which provides additional excitation channels to engineer scanning near-field microscopy and surface-enhanced spectroscopy.
\end{abstract}

\maketitle

\section{Introduction}
Condensed-matter theory describes the dielectric response of materials by collective dipole excitations that get induced by electric fields.\cite{Kittel2005,YuCardonaBook} This concept models optically active excitations of electrons, phonons, excitons, plasmons, and other quasiparticles;\cite{Basov2021} it explains the electrical susceptibility and dielectric function from the far infrared to the visible and soft X-ray regime.\cite{Kittel2005} The collective dipole excitations are usually divided into two distinct types: longitudinal --  parallel polarization and propagation directions -- and  transverse -- polarization and propagation directions are perpendicular. This separation is convenient in the Coulomb gauge, where only the transverse modes interact with the transverse electromagnetic field of propagating photons. Their interaction leads to the formation of hybrid light-matter states or polaritons for strong enough light-matter coupling.\cite{Huang19512,Hopfield1958} The longitudinal modes, on the other hand, are regarded as pure matter excitations and uncoupled to photons, for example, a longitudinal phonon is understood mainly as a mechanical wave.\cite{Huang19512,YuCardonaBook}  

Modeling dipolar longitudinal and transverse modes as fundamentally different leads to contradictions that have not been resolved so far.\cite{Cohen1955,Huang19512,Anderson1963,YuCardonaBook} Cohen and Keffer showed in a seminal paper that the longitudinal frequency $\omega_l$ is higher for vanishing wavevector $k$ than the corresponding transverse excitation $\omega_t$.\cite{Cohen1955} This well-known longitudinal-transverse (LT) splitting appears unphysical since the propagation direction becomes ill defined for $k=0$. Several proposals attempted to consolidate LT splitting with our general understanding of crystals,\cite{Anderson1963,YuCardonaBook,Cohen1955} most convincingly, by including retardation or light-matter coupling.\cite{Huang1951,Huang19512} When adding light-matter coupling to the crystal Hamiltonian, the upper transverse polariton at $\Gamma$ systematically coincides with the longitudinal dipole mode. While this solves the discrepancy for the eigenenergies, it also suggests a degeneracy between two different types of excitations, the transverse polaritons and the longitudinal matter mode. A systematic degeneracy appears impossible for the regimes of ultrastrong and deep strong light-matter coupling where the upper transverse polariton is almost exclusively composed of photons with vanishing contribution of matter excitations,\cite{DeLiberato2014,Mueller2020,Barros2021} while the longitudinal mode remains a pure matter state. \cite{Huang1951,Huang19512}

\begin{figure}
    \centering
    \includegraphics[width=0.95\columnwidth]{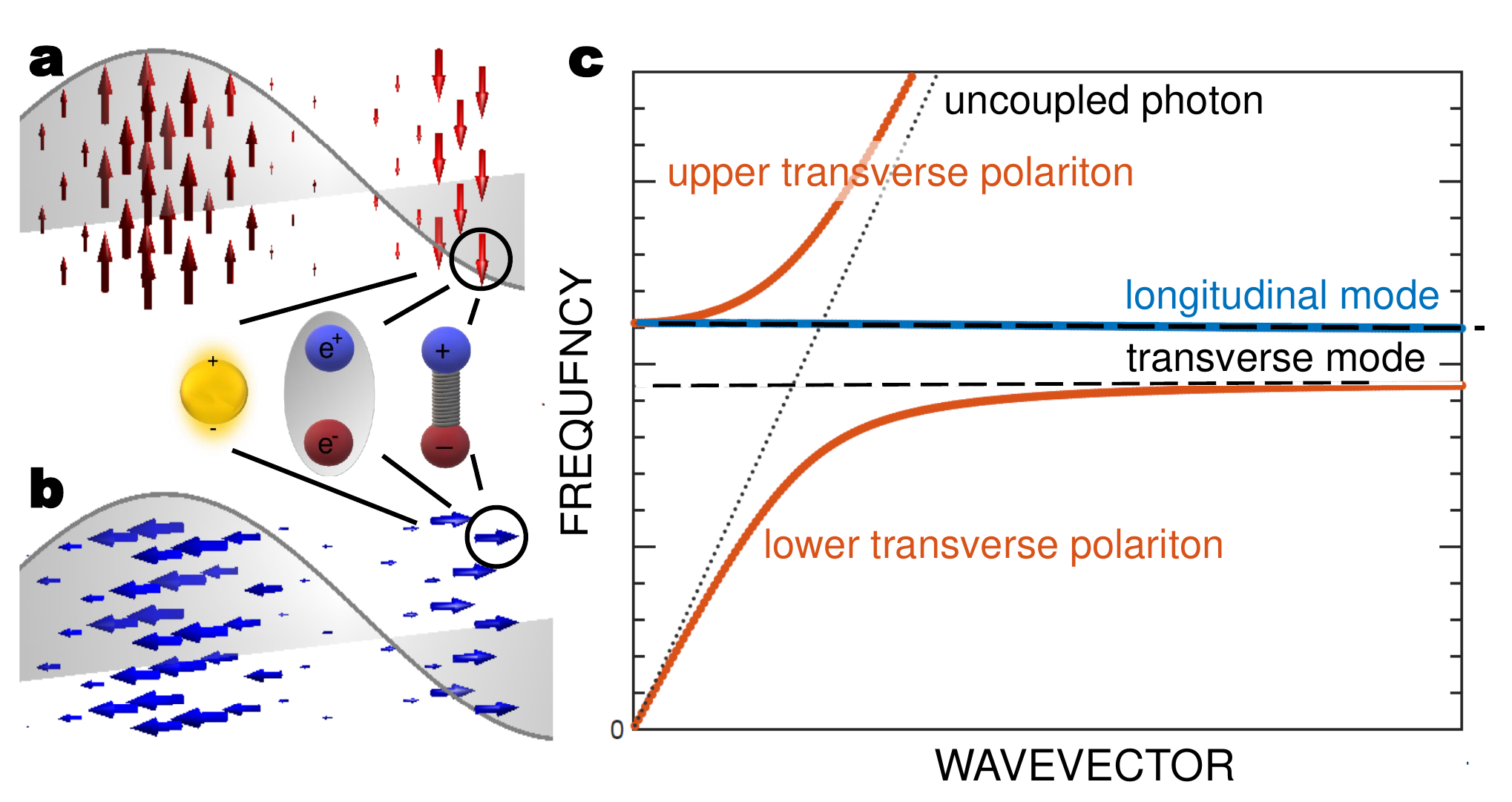}
    \caption{Induced dipoles and their dispersion. Sketch of the (a) transverse and (b) longitudinal collective modes. The dipoles may be associated with various material excitations, e.g.,  plasmons, excitons, and phonons. (c)  Polariton dispersion along the $\Gamma L$ direction of a face centered cubic lattice calculated in the Coulomb gauge. Dashed lines represent the longitudinal and transverse modes without light-matter coupling. }
    \label{fig:DipolePolariton}
\end{figure}

Here we show that in the Lorenz gauge the longitudinal and transverse dipole excitations are indistinguishable in the long wavelength limit, thus challenging the well established concept that only transverse excitations form polaritons. While the transverse modes continue to couple to transverse photons, the longitudinal modes couple to longitudinal and scalar photons making them also hybrid light-matter states. We show that the longitudinal and transverse polaritons are systematically degenerate at $k=0$; the apparent LT splitting arose from inconsistently disregarding the light-matter coupling to the transverse modes in the Coulomb gauge. Describing longitudinal and transverse modes as polaritons unifies our understanding of material excitations. It widens the choice of excitations for near-field optics and polariton devices by adding longitudinal polaritons as hybrid states of longitudinal and scalar photons with dipolar matter modes.This change of perspective also highlights the fundamental correspondence between light-matter and dipole-dipole interactions in crystalline systems.

This paper is organized as follows: we start by describing the effective quantum Hamiltonian for the lattice dipoles interacting with the electromagnetic field in the Lorenz gauge (Sect.~\ref{sec:Hamiltonian}). In Section \ref{sec:gBV} we discuss how to obtain the hybrid light-matter excitations through a generalized Bogoliubov transformation. In Sect.~\ref{sec:selfenergy} we describe how the dipole-dipole interaction can be described in the Lorenz gauge. The obtained dispersion relations and subsidiary conditions are presented in Sect.~\ref{sec:results} followed by a discussion (Sect.~\ref{sec:discussion}) of the broader implications of our results. Finally, we finish with some concluding remarks.

\section{Light-matter Hamiltonian in the Lorenz gauge\label{sec:Hamiltonian}} 

To describe materials excitations and their coupling to light we consider a lattice of isotropic bosonic dipole excitations with frequency $\omega_0$, Fig.~\ref{fig:DipolePolariton}(a) and (b), coupled to the free electromagnetic field. These excitations represent the localized contribution to crystal phonons, excitons, plasmons, spin waves, and so forth.\cite{Basov2021}  The coupled light-matter system is described by a semi-empirical Hamiltonian
\begin{equation}
    \mathcal{H}_\mathrm{LM}=\mathcal{H}_{mat}+\mathcal H_{EM}+\mathcal H_{int}.\label{eq:HLM}
\end{equation}
The matter part of the Hamiltonian is given by
\begin{equation}
\mathcal H_{mat}=\sum_{\mathbf k} \omega_0 b_{j,\mathbf k}^\dag b_{j,\mathbf k}, 
\end{equation}
where $b_{j,\mathbf k}^\dag=1/N \sum_{n} b_{j,n}^\dag(\mathbf r_n)\exp(i\mathbf k\cdot\mathbf r_n)$ and $b_{j,n}^\dag(\mathbf r_n)$ creates a dipole excitation of magnitude $h_{j,n}=(b_{j,n}+b_{j,n}^\dag)/\sqrt{2m\omega_0}$ at site $\mathbf r_n$ and polarized along the $j$ direction ($j=x,y,z$), $N$ is the total number of lattice sites. When treating $\mathcal H_\mathrm{LM}$ within the Coulomb gauge the dipole-dipole interaction between the localized excitations leads to longitudinal and transverse eigenstates, dashed lines in Fig.~\ref{fig:DipolePolariton}(c), that are understood as the pure crystal excitations. Their coupling to propagating photons results in the transverse lower and upper polariton, orange lines in Fig.~\ref{fig:DipolePolariton}(c), whereas the longitudinal mode remains unaffected by light-matter coupling, blue line in Fig.~\ref{fig:DipolePolariton}(c). 

To switch to the Lorenz gauge we write the photon in the covariant form ($c=\hbar=1$) as
\begin{equation}
A_\mu (\mathbf r,t)=\int d^3k \sqrt{\frac{1}{2 k}} \left[ a_\mu(\mathbf k) e^{i(\mathbf k \cdot \mathbf r-\omega t)}+a^\dag_\mu(\mathbf k) e^{i(\mathbf k \cdot \mathbf r-\omega t)}\right],    
\end{equation}
 where $a_j$ is the annihilation operator for the vector potential. Here $\mu$ runs between 0 and 3, 0 being related to the scalar potential $A_0(x_\mu)=\Phi(x_\mu)$. The creation and annihilation operators for the vector potential obey cannonical commutation relations, the scalar potential has some exquisite properties: $(a_0)^\dag=-a_0^\dag$ and $[a_0,a_0^\dag]=-1$. $a_0$ commutes with the other operators. Following the work of Babiker \cite{Babiker1982,Gupta1950,Bleuler1950}, the Hamiltonian for the free electromagnetic field in the Lorenz gauge is 
 \begin{equation}
    \mathcal H_{EM}^{\rm LG}=\int d^3k \sum_j k[a_j^\dag(\mathbf k) a_j(\mathbf k)-a_0^\dag(\mathbf k) a_0(\mathbf k)],  
\end{equation}
where $j$ runs over the three dimensions in space $x,y,z$. The light-matter interaction is 
\begin{equation}
    \mathcal H_{int}^{\rm LG}= \sum_{n} \frac{q^2}{2m}A_j^2(\mathbf r_n) + \frac{q}{m} \mathbf p_{n}\cdot\mathbf A(\mathbf r_n) -\rho(r_n)A_0(\mathbf r_n),\label{eq_Hint0}
\end{equation}
with $\mathbf p_{n}$ the momentum of the $n$-th dipole and $\rho(r)$ the charge distribution. Throughout this paper the vector notation $\mathbf V$ refers to vectors in 3D space. We define the charge distribution for each dipole as that obtained by two opposite point charges oscillating around their rest position at $\mathbf r_n$ with a total displacement of $\mathbf h_n$ - $\rho(\mathbf r_n)=\sum_n q\int d^3r [\delta(\mathbf r-\mathbf r_n \mathbf + \mathbf h_n/2)-\delta(\mathbf r-\mathbf r_n \mathbf - \mathbf h_n/2)]$. With this ansatz the charge density in reciprocal space is  \begin{equation}
\begin{split}
\mathbf \rho(\mathbf k)=&\int d^3r \rho(\mathbf r) \exp(i\mathbf k\cdot \mathbf r)=2iq\sum_n \sin(\mathbf k\cdot \mathbf h_n/2)\exp(i\mathbf k\cdot \mathbf r_n)\\\sim~&iq\sum_n (\mathbf k\cdot \mathbf h_n) \exp(i\mathbf k\cdot \mathbf r_n)=i \mathbf k \cdot q\mathbf h(\mathbf k),       \end{split}
\end{equation}
where we make the approximation that $|\mathbf h|$ is much smaller than the lattice spacing and define $\mathbf h(k)=\sum_n \mathbf h_n \exp(i\mathbf k\cdot\mathbf r_n)$. 

We then substitute $\rho(\mathbf k)$ into Eq.(\ref{eq_Hint0}) and expand $\mathbf A(\mathbf k)$, $\mathbf p(\mathbf k)$ and $\mathbf h(\mathbf k)$, in terms of creation and annihilation operators for the photon and dipole excitations. This yields the light-matter interaction part of the Hamiltonian as
\begin{equation}
\begin{split}
\mathcal H_{int}^\mathrm{LG}(\mathbf k)=&\sum_{j,j',\mathbf G}\frac{g^2\xi^2_{\mathbf G}}{\omega_0}(\hat e_{j}\cdot \hat e_{j'}) [(a_{j,\mathbf k}+a_{j,-\mathbf k}^\dag)(a^\dag_{j',\mathbf k+\mathbf G}+a_{j',-\mathbf k-\mathbf G})] +\\ 
+&ig\xi_{\mathbf G}(\hat e_{j}\cdot \hat e_{j'})[(b_{j,\mathbf k}-b_{j,-\mathbf k}^\dag)(a^\dag_{j',\mathbf k+\mathbf G}+a_{j',-\mathbf k-\mathbf G})]+\\
+&ig\frac{\hat e_j\cdot\hat n_{\mathbf k+\mathbf G} }{\xi_{\mathbf G}}[(b_{j,\mathbf k}-b_{j,-\mathbf k}^\dag)(a^\dag_{0,\mathbf k+\mathbf G}+a_{0,-\mathbf k-\mathbf G})],
\end{split}\label{eq:Hint}
\end{equation} 
where $\xi_{\mathbf G}=\sqrt{\omega_0/|\mathbf k+\mathbf G|}$, $\hat n_{\mathbf k+\mathbf G}$ is a unit vector along the $\mathbf k+\mathbf G$ direction,  $\hat e_j$ is the polarization direction for the photon or the dipole with wavevector $\mathbf k$ and $\hat e_{j'}$  for the states with $\mathbf k+\mathbf G$ and $V$ is the volume of the unit cell. $\mathbf G$ runs over all reciprocal lattice vectors. $g=\sqrt{q^2/4mV}$ is the light-matter coupling strength of the excitation, equal to the vaccuum Rabi frequency, and related to the reduced coupling strength $\eta=g/\omega_t=g/\sqrt{\omega_0^2-4g^2/3}$. 

In the Lorenz gauge $\mathcal H_{LM}$ is highly symmetric for transverse and longitudinal photon excitations. The matter $\mathcal H_{mat}$ and photon part $\mathcal H_{EM}^\mathrm{LG}$ as well as the coupling of the local dipoles to transverse and longitudinal photons, first two terms in $\mathcal H_{int}^\mathrm{LG}$, are independent of propagation direction. The longitudinal and transverse solutions are, therefore, identical when we neglect scalar photons, third term in Eq.~\eqref{eq:Hint}. The term is only present for longitudinal modes, \textit{i.e.}, if the polarization is parallel to the propagation direction, and vanishes for $k=0$. Solutions of $\mathcal H_{LM}$ are three-fold degenerate at $\Gamma$ and all eigenvectors are polaritons or coupled light-matter states, resolving the conflicting LT splitting.

\subsection{Subsidiary condition and dynamical matrix\label{sec:gBV} }

The Lorenz gauge introduces four degrees of freedom for the electromagnetic field (two transverse, one longitudinal, and one scalar), but two of these degrees of freedom must be removed when solving the Hamiltonian. This is done by imposing subsidiary conditions; excitations that violate the conditions remain unpopulated at all times.\cite{Gupta1950,Bleuler1950} For free photons in the Lorenz gauge, this condition is imposed by requiring that physical states are those for which
\begin{equation}
    \left[a_L(\mathbf k)-a_0(\mathbf k)\right]|\chi\rangle =0\label{eq:LorenzCondititon},
\end{equation}
where $L$ (for longitudinal) is the direction parallel to $\mathbf k$. For our system, we must impose a general form of the Lorenz gauge condition
\begin{equation}
(\nabla\cdot A-\partial_t A_0)^+|\chi\rangle=\left(kA_L-[A_0,\mathcal H]\right)^+|\chi\rangle=0.\label{eq_subs11}
\end{equation}
For this equation to be applicable, the operator $\left(kA_L-[A_0,\mathcal H]\right)$ must be written in terms of the creation and annihilation operators ($\psi_{\lambda\mathbf k}^\dag$, $\psi_{\lambda\mathbf k}^\dag$) of the coupled system. The $+$ superscript in Eq.(\ref{eq_subs11}) indicates that we take only the positive frequency (annihilation) part of the operators. $|\chi\rangle$ is a proper Fock state, for which the ground state $|\chi_0\rangle$ gets destroyed by all annihilation operators of the coupled system. We can define the contribution of the annihilation operator to the the Lorenz gauge condition operator
\begin{equation}
Z_\lambda(\mathbf k)=[\left(kA_L-[A_0,\mathcal H]\right),\psi_{\lambda,\mathbf k}^\dag],    
\end{equation}
see Appendix (\ref{app:subs}) for details, such that the subsidiary condition now reads
\begin{equation}
\sum_\lambda Z_\lambda(\mathbf k)\psi_{\lambda,\mathbf k}|\chi\rangle=0.    
\end{equation}
Since we are interested in linearly independent operators, we impose the stronger condition
\begin{equation}
Z_{\lambda}(\mathbf k)\psi_{\lambda,\mathbf k}|\chi\rangle=0,
\end{equation}
such that only operators with $Z_{\lambda}(\mathbf k)\rightarrow 0$ correspond to physically relevant excitations, \textit{i. e.} $\psi_{\lambda,\mathbf k}|\chi\rangle\neq0$. 

To describe hybrid light and matter excitations, we must search for a set of operators $\psi_{\lambda,\mathbf k}$ and $\psi_{\lambda,\mathbf k}^\dag$ that fulfill the following conditions
\begin{subequations}
\begin{align}
&i\frac{d}{dt}\psi_{\lambda,\mathbf k}=[\psi_{\lambda,\mathbf k},\mathcal H]=\omega_{\lambda,\mathbf k}\psi_{\lambda,\mathbf k}\label{eq_condition1},\\
&[\psi_{\lambda,\mathbf k},\psi_{\lambda',\mathbf k}^\dag]=\pm \delta_{\lambda\lambda'}, \text{and} \label{eq_condition2}\\
&Z_\lambda(\mathbf k) \psi_{\lambda,\mathbf k}|\chi\rangle=0\label{eq_condition3},
\end{align}
\end{subequations}
where $[~]$ is the commutator. Condition~(\ref{eq_condition1}) requires the solutions to be linearly independent and Condition~(\ref{eq_condition3}) imposes the Lorenz gauge. Condition~(\ref{eq_condition2}) ensures canonical (or anti-canonical) commutation relations. The negative sign is included because of the anti-Hermitian nature of the scalar potential.\cite{Cohen_tannouji}  Here, $|\chi\rangle$ is an appropriately defined Fock state and $Z_\lambda(\mathbf k)$  weights the contribution of $\psi_{\lambda,\mathbf k}$ to the subsidiary condition, see Appendix \ref{app:subs} for details.
Operators $\psi_{\lambda,\mathbf k}$ for which $Z_\lambda(\mathbf k)$ is non-zero require that $\psi_{\lambda,\mathbf k}|\chi\rangle$ vanishes at all times for all physically relevant states, thus being removed from the dynamics of the system. Ideally, two sets of operators will be eliminated through this process, thus effectively re-establishing the expected two degrees of freedom.

To ensure that Condition~(\ref{eq_condition1}) is fulfilled, we write the operators $\psi_{\lambda,\mathbf k}^\dag$ and $\psi_{\lambda,\mathbf k}$ as linear combinations of the bare light and matter creation and annihilation operators. This is known as a Bogoliubov-Valantin (BV) transformation. It is useful to write the Hamiltonian in terms of a Dynamical matrix,
\begin{equation}
    \mathcal H=\frac{1}{2}\sum_\mathbf{k}\Phi_\mathbf{k}^\dag \Sigma \mathcal D_{\mathbf k}
\Phi_\mathbf{k}.
\end{equation}
$\Phi_{k}=(\bar\phi_{k},~ \bar\phi_{-k}^\dag)$ is a $2N$-sized column vector consisting of the creation and annihilation operators for the $N$ bare excitations in the system. $\Sigma=\mathrm{diag}(1,~1,..,-1,-1,..)$ is the metric for the BV transformation.
For operators following the canonical commutation relations it is always possible to define the Bogoliubov vector $\Phi$ such that $[\Phi,\Phi^\dag]=\Sigma$ and the BV transformation $T_{\mathbf k}$ that changes the original set of operators $\Phi_\mathbf k$ to $\Psi_{\mathbf k}=T^{-1}_{\mathbf k}\Phi_{\mathbf k}$ and $T^{\dag}_{\mathbf k}\Sigma T_{\mathbf k}=\Sigma$,
\textit.{i. e.}, preserving the original canonical commutation relations $[\Psi, \Psi^\dag]=\Sigma$. If this BV transformation diagonalizes the Dynamical matrix, the new set of operators $\Psi_{\mathbf k}$ correspond to well defined and linearly independent creation and annihilation operators, following both Condition~(\ref{eq_condition1}) and (\ref{eq_condition2}).

The peculiar commutation relation of the scalar photon operator ($[\Psi, \Psi^\dag]=\Upsilon\neq\Sigma$) requires applying a generalized Bogoliubov-Valantin  (gBV) transformation to attempt diagonalizing the dynamical matrix. We here define the generalized BV transformation $\tilde T_{\mathbf k}$ that fulfills $\tilde T_{\mathbf k}^\dag \Sigma \tilde T_{\mathbf k}=\Upsilon=[\Phi, \Phi^\dag]$, thus preserving the commutation relations of the original operators. It is also necessary that $\tilde T_{\mathbf k}$ obeys $\tilde T_{\mathbf k} \Upsilon \tilde T_{\mathbf k}^\dag=\Sigma$. Ideally, this transformation also diagonalizes the Dynamical matrix, leading to linearly independent excitations which hold the original commutation relations. Unfortunately, this is sometimes impossible, and only a partial diagonalization is obtained. The  conditions for a given Dynamical matrix to be diagonalizable by a generalized BV transformation should be the subject of further pursuit. 

Even when the dynamical matrix cannot be fully diagonlized, there are instances on which we can still obtain physically meaningful solutions for the Hamiltonian. This occurs when only the linearly independent solutions obey the subsidiary conditions imposed by the Lorenz gauge. In such situations, the linearly dependent solutions are disregarded and the remaining solutions will describe the physical excitations of the system. 

\subsection{Dipole-dipole interaction \label{sec:selfenergy}}

The Lorenz gauge lacks a term for the dipole interaction of the crystal excitations and instead describes this interaction to occur through virtual longitudinal and scalar photons. This means that we have to include photons of all wavevector in our calculation to properly describe the dipole-dipole coupling, although we only calculate the crystal excitations within the first Brillouin zone. We will term the effect of longitudinal and scalar photons outside the first Brillouin zone onto the polariton dispersion as Umklapp contributions. They are vital for correctly describing collective crystal excitations. 

We add Umklapp contributions by the longitudinal and scalar photons as a self-energy correction to the dynamical matrix in the 1st Brillouin zone. Umklapp contributions due to transverse photons, in contrast, may be discarded. To obtain the self-energy correction, we organize the full dynamical matrix for a given $\mathbf k$ into the form
\begin{equation}
D_{\mathbf k}^\mathrm{full}=\begin{pmatrix}
D_{\mathbf k}^{0} & \kappa_{\mathbf k} \\   
\bar\kappa_{\mathbf k} & D_{\mathbf k}^{U} 
\end{pmatrix}.    
\end{equation}
$D_{\mathbf k}^{0}$ is a block diagonal combination of the transverse and longitudinal dynamical matrices (considered to be independent), $D_{\mathbf k}^{U}$ is the dynamical matrix for the Umklapp terms, and $\kappa$ is the interaction between $\mathbf G=0$ and $\mathbf G\neq0$ terms. Note that $\bar \kappa \neq \kappa^\dag$ since the dynamical matrix is not Hermitian.

We obtain the Green's function for the $\mathbf G=0$ part as $\mathcal G^{\mathrm 0}_{\mathbf k} = (D_{\mathbf k}^{0}-\Delta_{\mathbf k}(\omega)-\omega I)^{-1}$, where $\Delta_{\mathbf k}(\omega)=\bar\kappa_{\mathbf k} \mathcal G_{\mathbf k}^{U}(\omega) \kappa_{\mathbf k}$ is the self-energy correction due to the interaction with the Umklapp excitations. Here  $\mathcal G_{\mathbf k}^{U}(\omega)=(D_{\mathbf k}^{U}-\omega I)^{-1}$ is the Green's function for the Umklapp terms. For a cubic lattice, on which the three axis directions can be interchanged, the self-energy correction, see Appendix \ref{app:DLG}, is expressed as 
\begin{equation}
\delta_{\mathbf k}(\omega)=\frac{q^2}{4mV\omega_0}\sum_{\mathbf G \neq0}2(\hat k\cdot\hat n_{\mathbf k+\mathbf G})^2 \frac{\omega_0^2-|\mathbf k +\mathbf G|^2}{|\mathbf k +\mathbf G|^2-\omega^2}=\frac{2}{3}\frac{g^2}{\omega_0}\sum_{\mathbf G \neq0} \frac{\omega_0^2-|\mathbf k +\mathbf G|^2}{|\mathbf k +\mathbf G|^2-\omega^2}, \label{eq:deltaT}
\end{equation}
Within this approximation the self-energy correction is the same for both longitudinal and transverse dipole modes. We now note that if we add and subtract the $\mathbf G=0$ contribution to $\Delta_\mathbf{k}(\omega)$ we get 
\begin{equation}
\delta_{\mathbf k}(\omega)=\frac{2g^2}{3\omega_0} \left(\frac{\omega_0^2 - k^2}{\omega^2-k^2}\right)+\frac{2g^2}{3\omega_0}\sum_{\mathbf G} \frac{\omega^2_0-|\mathbf k +\mathbf G|^2}{|\mathbf k +\mathbf G|^2-\omega^2}.
\end{equation}
At $k=0$, the second term in the right-hand side
corresponds to the self-energy of the isolated dipoles interacting with the free-electromagnetic field. This term diverges and must be removed by renormalization. Here we ignore it to obtain the final self-energy correction due to the dipole-dipole interaction
\begin{equation}
\delta_{\mathbf k}(\omega)=\frac{2}{3}\frac{g^2}{\omega_0}\left(\frac{\omega^2+k^2}{\omega_0^2-k^2}\right). \label{eq:deltak}
\end{equation}
For $k\rightarrow 0$, the self-energy can be assumed to be constant and evaluated at $\omega=\omega_0$ leading to $\delta_0(\omega_0)=\frac{2}{3}\frac{g^2}{\omega_0}$. In the Lorenz gauge, the connection between light-matter and dipole-dipole interactions becomes explicit. The dipole-dipole interaction is described in terms of virtual photon excitations and is parameterized by the light-matter coupling constant $g$. Any system with a strong dipole-dipole interaction should also experience strong light-matter coupling effects, and vice-versa.  

With the self energy correction we are able to include the Umklapp processes in the dynamical matrix and thus describe the interaction between dipoles within the Lorenz gauge. It turns out that the dynamical matrix may be separated into a transverse (denoted by superscript T in the following equations) and a longitudinal (L) part
\begin{equation}\label{eq:DLT}
D^{T,L}_{\mathbf k}=\begin{pmatrix} \alpha^{T,L}_{\mathbf k} & \gamma^{T,L}_{\mathbf k}\\
-\gamma^{T,L\dag}_{\mathbf k} & -\alpha^{T,L\ast}_{\mathbf k}    
\end{pmatrix}, 
\end{equation}
with
\begin{equation}\label{eq:alpgammaT}
\alpha^T_{\mathbf k}=\begin{pmatrix}
\omega_0-\frac{2}{3}\frac{g^2}{\omega_0} & ig\xi\\
-ig\xi & k+\Xi  
\end{pmatrix},
\gamma^T_{\mathbf k}=\begin{pmatrix}
-\frac{2}{3}\frac{g^2}{\omega_0} & ig\xi\\
ig\xi & \Xi 
\end{pmatrix},
\end{equation}
and
\begin{equation}\label{eq:alpgammaL}
\alpha^L_{\mathbf k}=\begin{pmatrix}
\omega_0-\frac{2}{3}\frac{g^2}{\omega_0} & ig\xi & i g/\xi\\
-ig\xi & k+\Xi & 0 \\
-ig/\xi & 0 & - k  
\end{pmatrix},
\gamma^L_{\mathbf k}=\begin{pmatrix}
-\frac{2}{3}\frac{g^2}{\omega_0} & ig\xi & i g/\xi\\
ig\xi & \Xi & 0 \\
ig/\xi & 0 & 0 
\end{pmatrix},
\end{equation}
where $\omega_0$ is the frequency of the dipole oscillations and $\Xi=2g^2|\xi|^2/\omega_0$. With $\xi=\xi_{\mathbf G=0}=\sqrt{\omega_0/k}$. We will attempt to diagonalize the dynamical matrix in the following section. 

\section{Polaritons in the Lorenz gauge\label{sec:results}}

\begin{figure*}
    \centering
    \includegraphics[width=8.1cm]{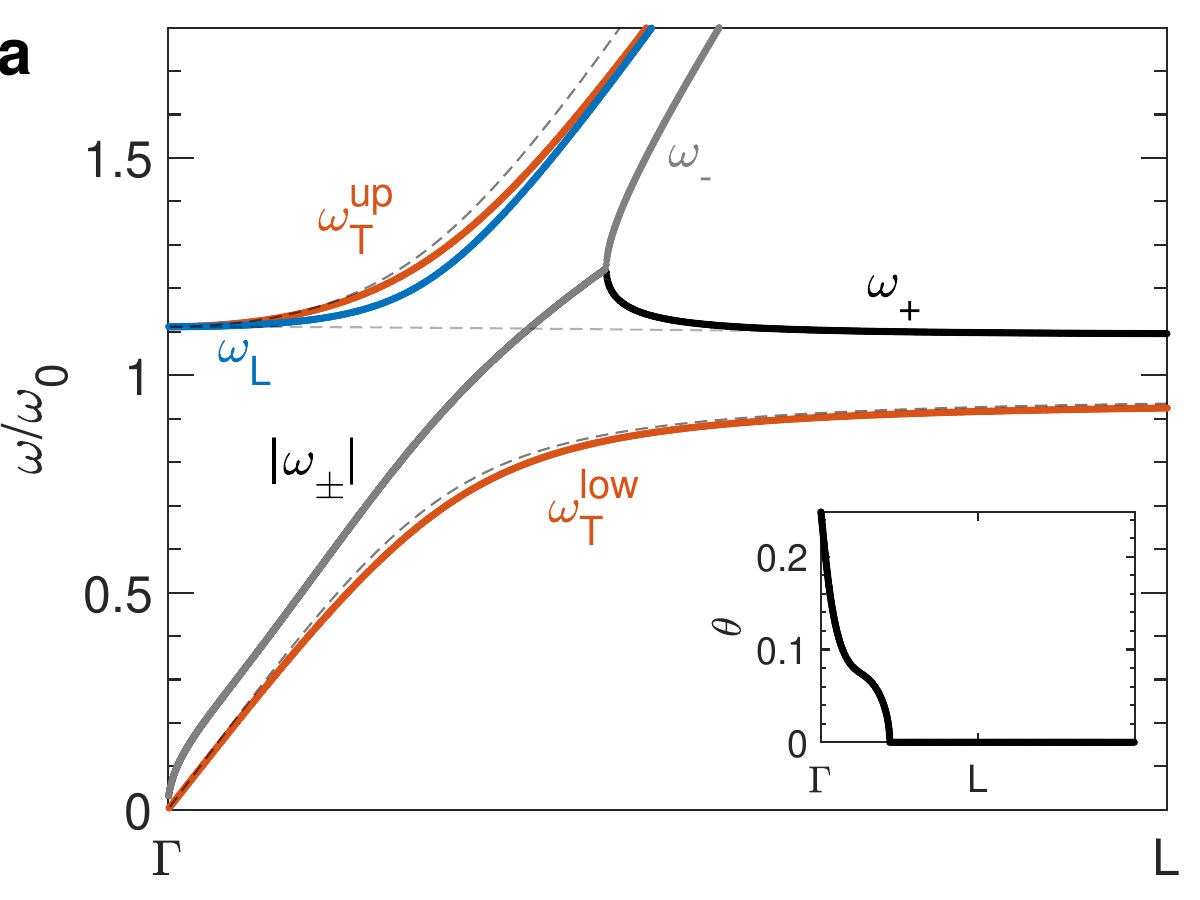}
    \includegraphics[width=8.1cm]{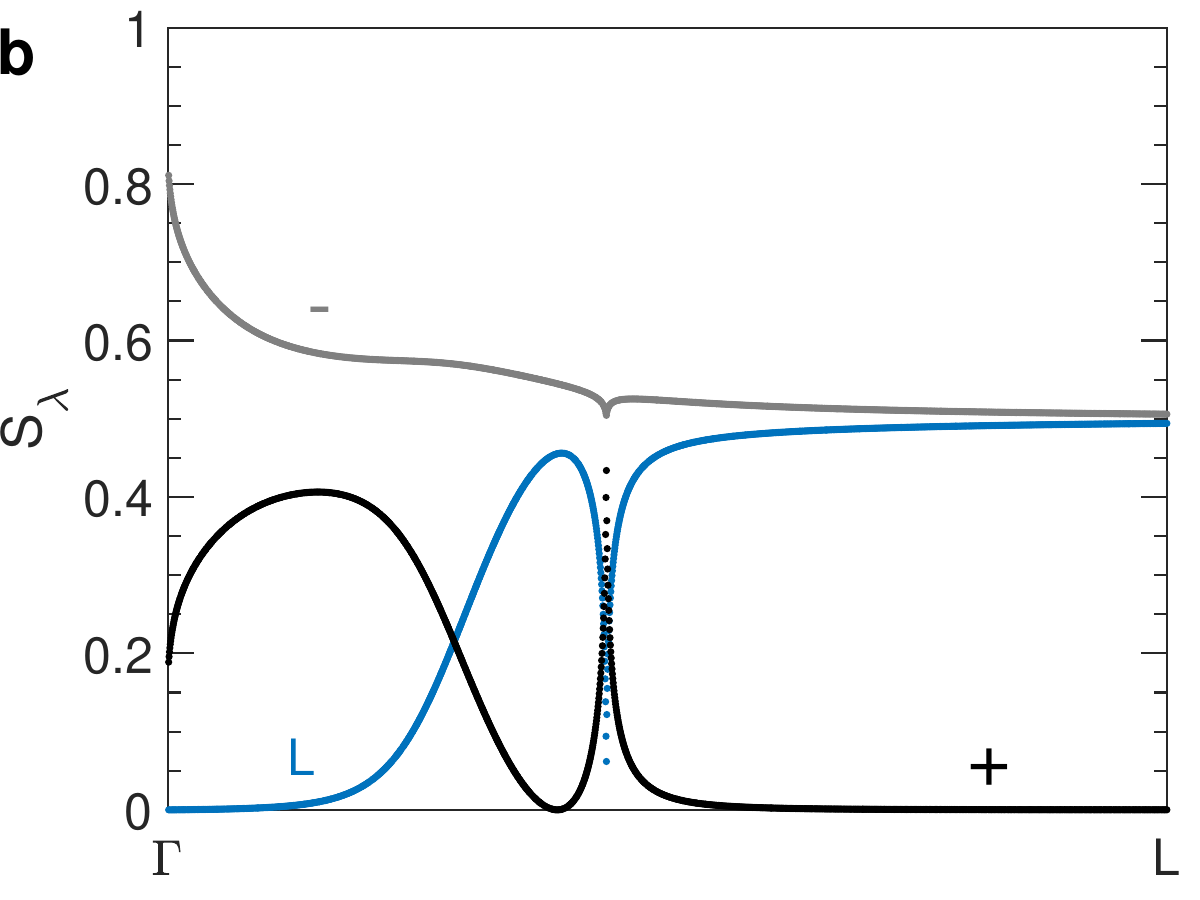}
    \caption{
    (a) Dispersion in the Lorenz gauge of the transverse (red) polariton, and longitudinal (blue, black and gray)  solutions, see labels. The inset shows the phase ($\theta$) of the longitudinal modes for a diagonalization by a similarity transformation, see text for details.   The dashed lines  correspond to the solutions in the Coulomb gauge. (b) Relative contribution, $S_\lambda=|Z_\lambda|^2/\sum |Z_\lambda|^2$, of the longitudinal modes to the subsidiary condition. Only states with $Z_{\lambda}\approx0$ are considered as physical states. $\eta=0.26$, $a\omega_0=\pi/3$.   \label{fig:wk}}
\end{figure*}

With the transverse and longitudinal parts of dynamical matrix we may obtain the excitations of the coupled light-matter systems within the Lorenz gauge. The transverse part of the dynamical matrix is diagonalized by a regular BV transformation and yields two physically acceptable hybrid light-matter states or polaritons, full red lines in Fig.~\ref{fig:wk}(a). They are close in energy to the transverse solutions in the Coulomb gauge, dashed lines. The remaining discrepancies arise from disregarding the wavevector dependence of the self-energy correction.

The longitudinal part of the dynamical matrix involves the interaction of the dipoles with longitudinal and scalar photons and cannot be solved trivially. To circumvent this, we first diagonalized it numerically by conjugation to obtain the eigenfrequencies $\omega_{L,\mathbf k}$, $\omega_{+,\mathbf k}$, and $\omega_{-,\mathbf k}$ in Fig.~\ref{fig:wk}(a). $\omega_{L,\mathbf k}$ is real, blue line in Fig.~\ref{fig:wk}(a), while the set ($\omega_{+,\mathbf k}$, $\omega_{-,\mathbf k}$) has the characteristics of non-Hermitian parity-time symmetric Hamiltonians, such as the presence of an exceptional point $k_{\mathrm{EP}}$. \cite{Minganti2019, Ozdemir2019} The eigenspectrum includes two complex frequencies $\omega_{+,\mathbf k}=\omega_{-,\mathbf k}^\dag=\omega_{\pm}e^{\pm i\theta}$ for $k<k_{\mathrm{EP}}$ and two real frequencies $\omega_{+,\mathbf k}\neq\omega_{-,\mathbf k}$ for $k>k_{\mathrm{EP}}$, grey and black lines in Fig.~\ref{fig:wk}(a). Although diagonalizing numerically by conjugation gave us the eigenfrequencies, it does not allow for describing the solutions as creation and annihilation operators, which we need to identify the physically relevant states. For this, we apply the proposed generalized BV transformation that partially diagonalizes the dynamical matrix, see Appendix \ref{app:diag}, leading to one fully independent solution with eigenenergy $\omega_{L,\mathbf k}$ and two coupled solutions, $\psi_{+,\mathbf k}$ and $\psi_{-,\mathbf k}$. The operator $\psi_L$ obeys the three Conditions~(\ref{eq_condition3}) close to the $\Gamma$ point, as $Z_{\mathbf k,L}$ vanishes as $k^2$, blue line in Fig.~\ref{fig:wk}(b).  It is a composition of bare longitudinal dipoles and longitudinal and scalar photons and represents the physical excitations of the system with independent dynamics. In contrast, $Z_{\mathbf k,+}$ and $Z_{\mathbf k,-}$ are non-zero close to $\Gamma$ and do not correspond to physical excitations, Fig.~\ref{fig:wk}(b), \textit{i. e.}, these linearly dependent excitations must remain unpopulated in any physically relevant state.   

Although our model was motivated by understanding material excitation close to $\Gamma$, we find it illuminating to consider the dispersion throughout the Brillouin zone. With increasing $k$ we reach a critical point at $k=k_\mathrm{EP}$ that separates the long-wavelength behavior close to $\Gamma$ from the short-wavelength behavior in the rest of the Brillouin zone, Fig.~\ref{fig:wk}(a). Exceptional points occur in the solutions of non-Hermitian Hamiltonians.\cite{Minganti2019,Ozdemir2019} None of the excitations  are physically relevant at $k_\mathrm{EP}$, Fig.~\ref{fig:wk}, because $Z_L,Z_+,$ and $Z_-$ are simultaneously different from zero, which means that our choice for fixing the gauge is inappropriate to describe the system at these $k$. The Lorenz gauge is not completely fixed, because the vector potentials may be shifted by $A_L\rightarrow A_L+\nabla f$ and $A_0\rightarrow A_0-\partial f/\partial t$ for a wide class of functions $f$. There should be a gauge fixing condition that results in physically relevant states for all wavevectors, which calls for further work to fully describe condensed-matter excitations in the Lorenz gauge. For $k>k_\mathrm{EP}$ and the coupling parameters used here, the contribution of the $\psi_+$ term ($Z_{+}$) decreases and becomes zero around $2/3\Gamma L$, whereas $Z_L$ and $Z_-$ remain finite. For large $k$, $Z_+$ describes the one physically relevant state of the system. The  flat longitudinal band in Fig.~\ref{fig:wk}(a), black line, close to the zone edge coincides with the dispersion of the longitudinal mode in the Coulomb gauge, dashed lines.
\begin{figure*}
    \centering
    \includegraphics[width=8cm]{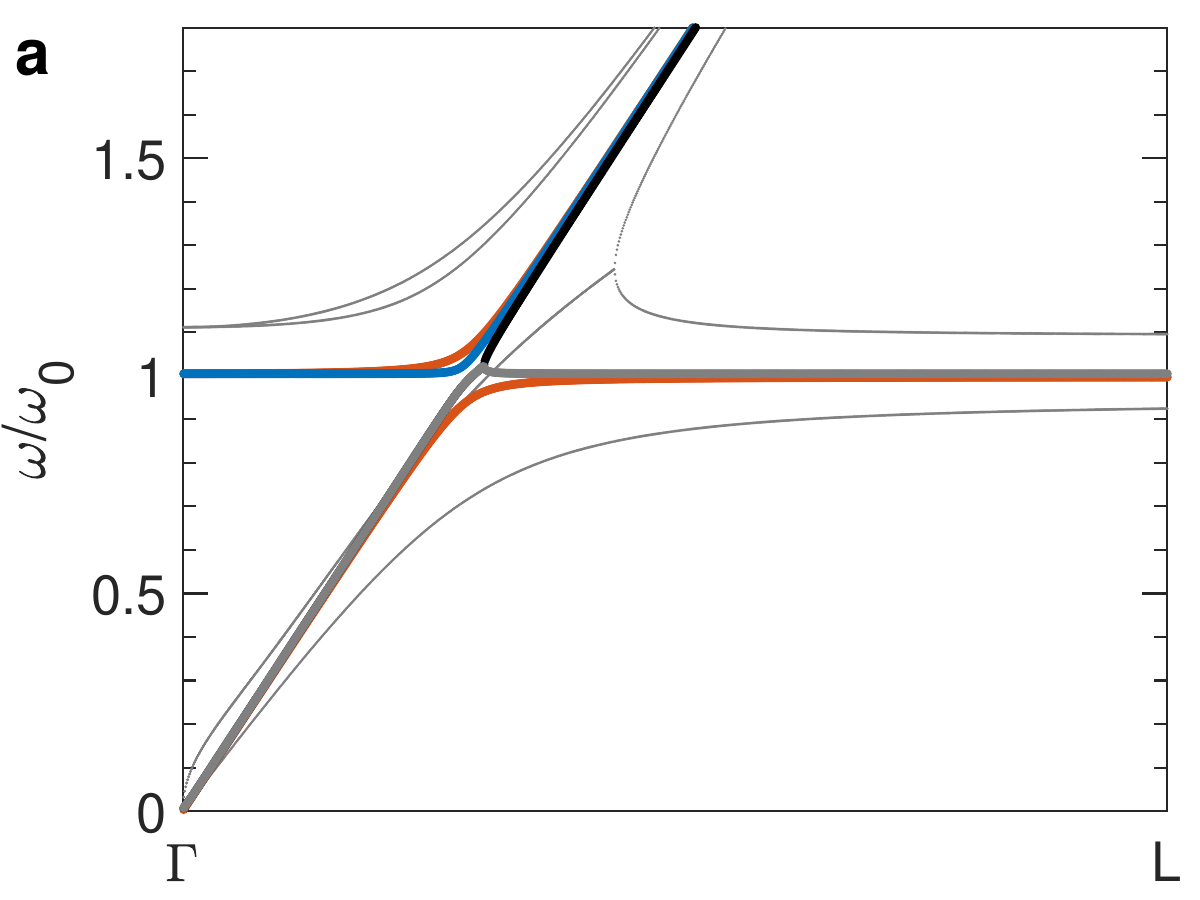}
    \includegraphics[width=8cm]{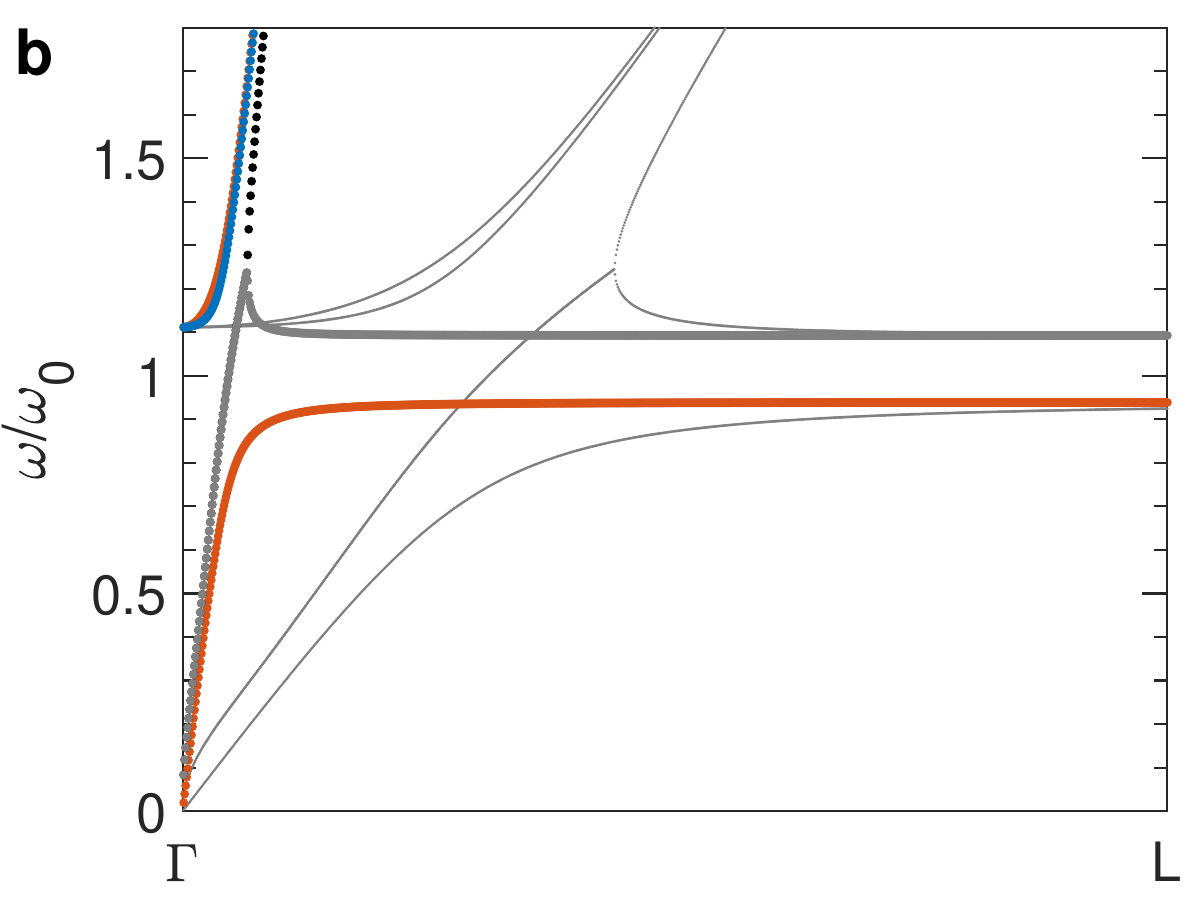}
    \caption{Polariton dispersion for varying lenght and interaction conditions. \textbf{a} $\eta=0.05$ and $a\omega_0=\pi/3$ and \textbf{b} $\eta=0.26$ and $a\omega_0=\pi/20$. For comparison, the dispersion of Fig.~\ref{fig:wk} ($g/\omega_0=0.25$ and $a\omega_0=\pi/3$) is shown in thin grey lines. Colors as in Fig.~\ref{fig:wk}    }
    \label{fig:different-parameters}
\end{figure*} 

Decreasing the strength of the light-matter coupling $g$ shifts $k_\mathrm{EP}$  closer to  $k=\omega_0$, where the photon and the dipole excitation have the same energy,  Fig.~\ref{fig:different-parameters}(a);
increasing $\eta$ moves $k_\mathrm{EP}$ to larger $k$ so that it may cross the zone edge. The $\mathbf G\neq 0$ photons then interact resonantly with the excitations and need to be considered explicitly. Increasing $g$ also moves the exceptional point up in energy and increases the (Rabi) splitting between the upper and lower transverse polaritons, Fig.~\ref{fig:different-parameters}(a). The second parameter governing the polariton dispersion is the lattice constant $a$ that moves the crossing point $k=\omega_0$ without affecting the Rabi splitting $\Omega_R$. The lattice constant of natural crystals are on the order of $a\sim 10^{-1}$\,m and $a\cdot\omega_0<<1$ as for the simulation in Fig.~\ref{fig:different-parameters}(b). In this case, both polariton branches are mainly matter like close to $\omega_0$, because the lower transverse polariton and the short-wavelength longitudinal polariton ($k>k_\mathrm{EP}$) are dominated by matter excitations.\cite{DeLiberato2014,Mueller2020} Metamaterials and artificial supercrystals have $a\approx 10-1000\,$nm meaning that the crossing point $k=\omega_0$ occurs in the middle of the Brillouin zone, Fig.~\ref{fig:wk}(a). The photon-like part of the longitudinal mode will be more accessible in such engineered structures.

\section{Discussion \label{sec:discussion}}

\begin{figure*}
    \centering
    \includegraphics[width=\textwidth]{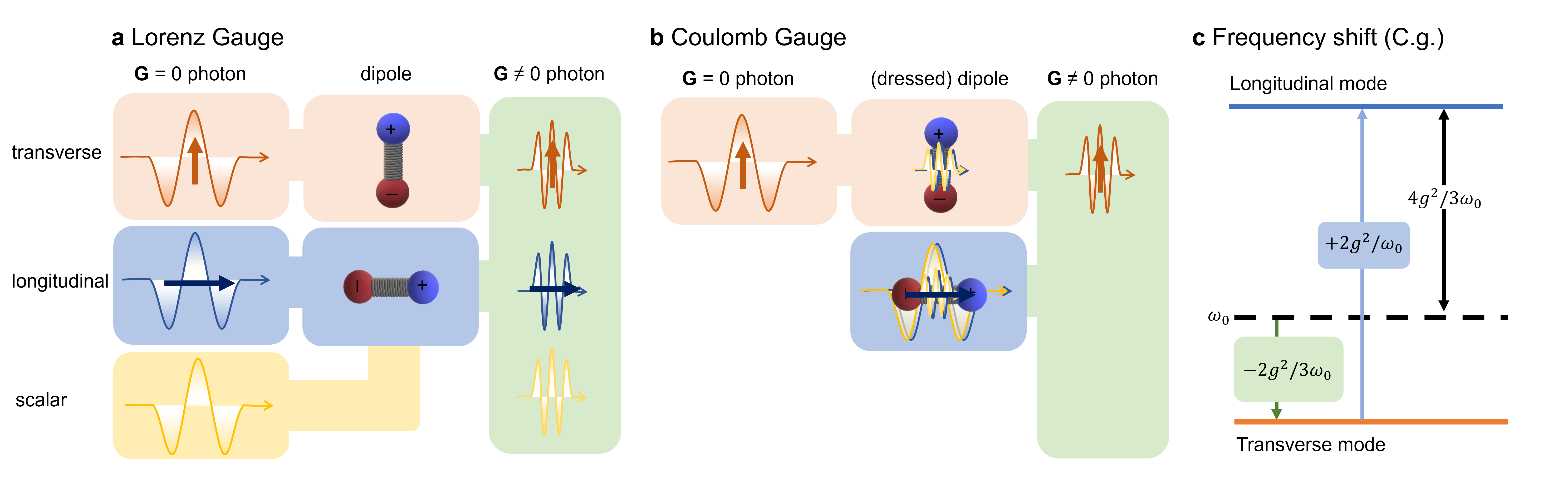}
    \caption{Light-matter coupling in the Lorenz and Coulomb gauge. \textbf{a} In the Lorenz gauge the long-wavelength transverse photons couple to transverse dipoles ($\mathbf{G}=0$, orange), long-wavelength longitudinal and scalar photons  couple to longitudinal dipoles ($\mathbf{G}=0$, blue). All short wavelength photons couple to all dipoles ($\mathbf{G}\neq0$, green). All components of light-matter coupling are treated explicitly; the dipoles remain undressed. \textbf{b} In the Coulomb gauge the long-and short-wavelength longitudinal and scalar photons dress the longitudinal dipoles. They become inseparable from the longitudinal mode that always has a photonic contribution (blue). The longitudinal and scalar short wavelength photons dress the transverse dipoles ($\mathbf{G}\neq0$, green). The long-wavelength transverse photons couple to the transverse dipoles. The short wavelength transverse photons couple to both dipoles. Only the coupling to the transverse modes is treated explicitly. 
\textbf{c} Energy shifts in the Coulomb gauge due to the dressing by photons. Transverse and longitudinal modes shift down due to the inclusion of the short-wavelength photons (green). The longitudinal modes shift up through the long-wavelength longitudinal and scalar photons (blue). The transverse mode remains finite and at lower frequency when neglecting the coupling to the transverse photons.}
    \label{fig:sketch}
\end{figure*}

It is fascinating to examine how the description of materials excitations in the Lorenz and Coulomb gauge shape our understanding of matter, Fig.~\ref{fig:sketch}. It may even lead to apparent gauge ambiguities similar to the ones observed in cavity electrodynamics for inconsistent approximations in the various gauges. \cite{DiStefano2019,Rouse2021} In the Lorenz gauge the dipole excitations remain undressed. Their coupling to light is treated explicitly for all photons. All collective eigenstates are polaritons; the coupling of the longitudinal and transverse photons to matter is completely symmetric. In the Coulomb gauge the dynamical parts of some photons get infused with the excitations creating dressed collective states that contain the photons in an inseparable way. All dipole excitations get dressed by the short-wavelength scalar and longitudinal photons, dressed dipole in Fig.~\ref{fig:sketch}(b), reducing the frequency of the single dipole $\omega_0$ to the collective frequency, Fig.~\ref{fig:sketch}(c). The longitudinal collective excitations are then dressed by the long-wavelength longitudinal and scalar photons increasing the longitudinal frequency to a net blue shift $+4g^2/3\omega_0$ compared to $\omega_0$, see Fig.~\ref{fig:sketch}(b) and (c). 
This last step gets introduced in materials modeling as an electric field produced by the longitudinal excitations that results in an additional frequency term, e.g., the Born effective charge for phonons in ionic crystals. The transverse photons remain separate from the transverse dipole excitations in the Coulomb gauge. Only this coupling is made explicit in the so-called light-matter interaction term that may or may not be included in the calculation. The appearance of a transverse mode with non-zero frequency below $\omega_0$ at $\Gamma$ and the resulting LT splitting is a consequence of neglecting one part of light-matter interaction, namely the separable part related to the transverse mode, while including the inseparable part in the longitudinal mode that was incorrectly understood as a matter-only excitation. 

For small wavevectors the light-matter Hamiltonian within the Lorenz gauge leads to one physically relevant longitudinal polariton. It is degenerate with the two transverse upper polaritons at $k=0$. The LT degeneracy is a general result in solids, it occurs independently of crystal structure and the magnitude of the light-matter coupling. We only assumed that the individual dipole excitations to be isotropic and the longitudinal and transverse modes to be polarized along the same direction. We also observed it systematically in binary  lattices
of plasmonic nanopaticles with two independent dipole excitations of different frequencies.\cite{Epishin2023} The LT degeneracy, however, does not occur in low-dimensional systems: We can describe a free-standing two-dimensional layer as a three-dimensional crystal with one lattice vector taken to infinity or infinitely small reciprocal lattice vector. The $\mathbf{G}\neq0$ terms remain relevant even for vanishing $k$ resulting in different self-energy corrections for the transverse and longitudinal modes, which lifts the LT degeneracy.

\begin{figure}
    \centering
    \includegraphics[width=8.5cm]{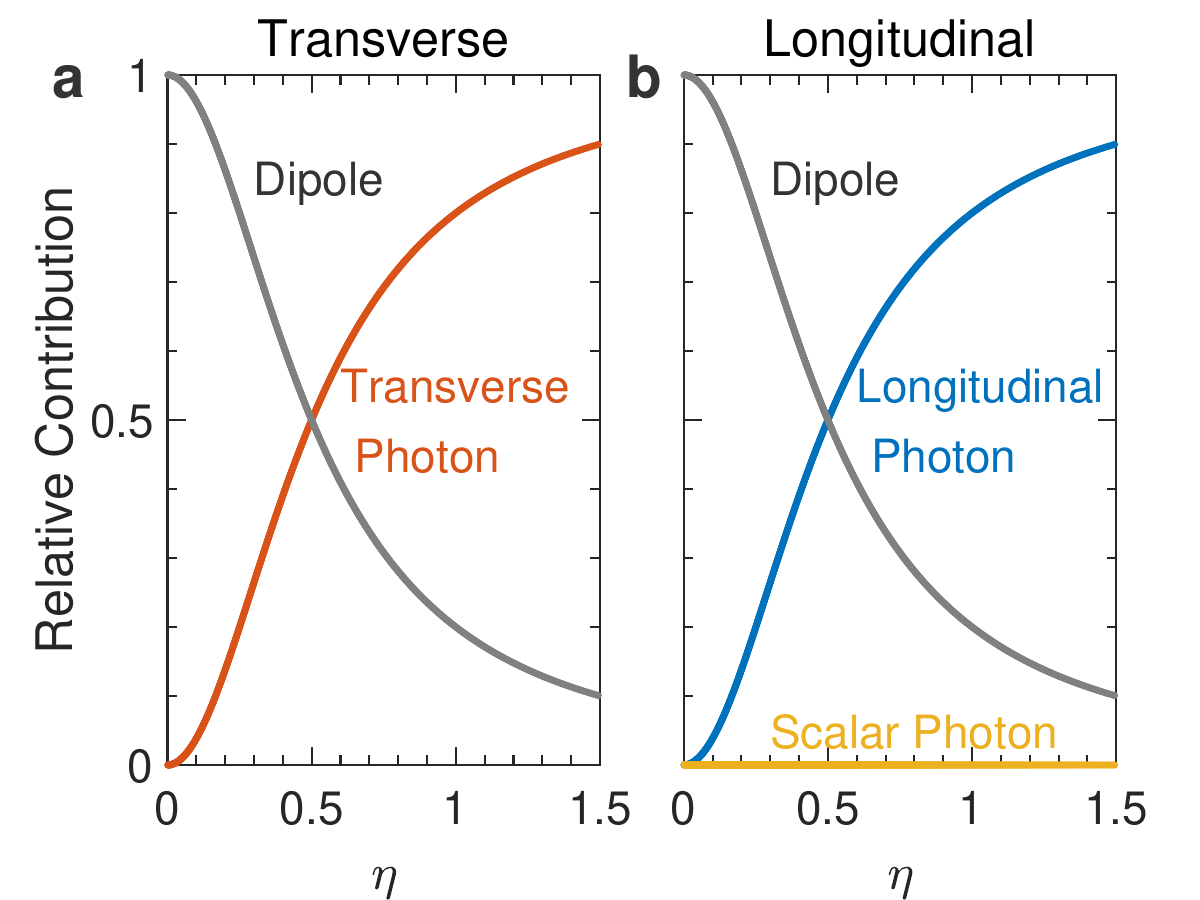}
    \caption{Relative contribution of the bare excitations to the (a) transverse, and  (b) longitudinal polariton near $k=0$ as a function of the reduced light-matter coupling strength $\eta$. For large $\eta$ the upper polaritons are almost exclusively composed of photons.  $k=0.01\pi/a$ and $a\omega_0=\pi/20$.}
    \label{fig:weights}
\end{figure}

For three-dimensional crystals, the Lorenz gauge resolves the vexing degeneracy between the transverse polariton and the longitdudinal matter state, i.e., two excitations of different character, that arose in the Coulomb gauge.\cite{Huang1951,Huang19512} 
Both  transverse and  longitudinal branches are hybridized states that are composed of  bare light and matter excitations as illustrated in Fig.~\ref{fig:sketch}. In Figure~\ref{fig:weights} we decomposed the upper transverse [panel (a)] and the longitudinal polariton [panel (b)]  into their contributions within the long wavelength limit. De Liberato\cite{DeLiberato2014} predicted the upper transverse polariton to become more and more photon like with increasing light-matter coupling as shown in Fig.~\ref{fig:weights}(a), which was later confirmed experimentally by Mueller~\textit{et al.}\cite{Mueller2020} In Fig.~\ref{fig:weights}(b) we calculated the corresponding decomposition for the longitudinal polariton into dipole or matter excitation (grey) as well as  longitudinal (blue) and scalar (yellow) photons. The figure shows that the behavior of the two types of excitations is identical close to the Brillouin zone center, as expected from the symmetriy of the Hamiltonian in this limit. 

With increasing reduced coupling the photon contribution to the longitudinal polariton becomes comparable to the matter contribution ($\eta\gtrsim 0.5$) and, finally, dominates for $\eta>1$. For the transverse modes, this effect leads to remarkable consequences such as the breakdown of the Purcell effect for particularly strong light-matter coupling.\cite{DeLiberato2014,Mueller2020,Barros2021} Corresponding effects may be expected also for the longitudinal modes.

\section{Conclusions\label{sec:conclusions}}

In conclusion, we presented a treatment of collective dipolar excitations in crystals within the Lorenz gauge, as an alternative to the commonly employed Coulomb gauge. The Lorenz gauge predicts a systematic degeneracy in energy and character of the transverse and longitudinal crystals excitations for vanishing wave vector resolving the long-standing puzzle of longitudinal-transverse splitting. It also explicitly highlights the contribution of longitudinal and scalar photons to crystal excitations and, thereby, shows that optically active transverse and longitudinal excitations in crystals are, in fact, coupled light-matter states.  The Lorenz gauge describes polaritons in the long-wavelength limit without using Ewald summation techniques nor producing discontinuities. It may also lead to an unambiguous description of static homogeneous polarization in crystals without resorting to Berry curvatures and phases, as required for the modern theory of polarization.\cite{Spaldin2012} With our current implementation, these insights come at the price of  exceptional points in the dispersion relation, where the excitations are ill defined, calling for future work to fully understand light-matter coupling in three dimensional crystals. 

By describing the longitudinal excitations in materials as hybrid light-matter states we allow for a deeper understanding of their physical properties within a quantum mechanical framework. This point of view suggests an important shift in paradigm. Inspired by the role virtual longitudinal and scalar photons play in describing the interaction of charges in vaccum,\cite{Cohen_tannouji} virtual polariton excitations in the Lorentz gauge may be instrumental to develop an alternative quantum description of the screened Coulomb interaction in materials. Longitudinal polaritons in the ultra-strong and deep-strong coupling regimes should experience similar physical effects as predicted for transverse polaritons, \textit{e. g.} a decoupling of light and matter, vanishing Purcell effect for near-field excitations, squeezed photons and virtual excitations in the ground state.\cite{DeLiberato2014,Kockum2019,FornDiaz2019} Most experiments on ultrastrong coupling focused on transverse excitations. Recent developments in tayloring plasmonic and photonic structures for novel energy and spatial scales will widen the path for additionally studying longitudinal polaritonic excitations, unravelling new related phenomena and engineering longitudinal and scalar photon modes in materials. 

\section{Acknowledgements}
The authors thank C.\,Thomsen for a critical reading of the manuscript and useful discussions. This work was supported by the European Research Council (ERC) under grant DarkSERS-772\,108, the German Science Foundation (DFG) under grant 504\,656\,879, the Center for International Collaboration (CIC), the Berlin Center for Global Engagement (BCGE), and the SupraFAB Research Center at Freie Universit\"at Berlin. E.B.B. acknowledges support from FUNCAP (PRONEX PR2-0101-00006.01.00/15), CNPq, and CAPES.

\appendix

\section{Dynamical matrices in the Lorenz gauge \label{app:DLG}}

Here we show how we obtain the independent dynamical matrices the longitudinal and transverse excitations in the first Brillouin zone. Specifically, we show how the interaction of the dipoles with the scalar and longitudinal Umklapp photons is described in terms of a self-energy correction which gives rise to the dipole-dipole interaction in Coulomb gauge. 

If we disregard photon-photon interaction for Umklapp photons, the dynamical matrix for the light-matter Hamiltonian can be written as
\begin{equation}
D_{\mathbf k}^\mathrm{full}=\begin{pmatrix}
D_{\mathbf k}^{0} & \kappa_{\mathbf k} \\   
\bar\kappa_{\mathbf k} & D_{\mathbf k}^{U} 
\end{pmatrix}=\begin{pmatrix}
\alpha_{0} & \gamma_0 & \kappa_{1} & \kappa_2  \\   
-\gamma_0^\dag & -\alpha_0^t & -\kappa_2 & -\kappa_1^\ast \\
\kappa_1^\dag & \kappa_2^\dag & \alpha_V & \gamma_V \\
-\kappa_2^\dag & \kappa_1^t & -\gamma_V^\dag & -\alpha_V^\ast
\end{pmatrix},    
\end{equation}
where 
\begin{equation}
\alpha_0=\begin{pmatrix}
\omega_0 I_3 & ig\xi_0 I_3 & ig\xi_0 k/\omega_0 \\
-ig\xi_0 I_3 &  kI_3+2g^2\xi_0^2/\omega_0 I_3 & 0\\
-ig\xi_0 k/\omega_0 & 0 & -k 
\end{pmatrix}   
\end{equation}
and
\begin{equation}
\gamma_0=\begin{pmatrix}
0 I_3 & ig\xi_0 I_3 & ig/\xi_0 \\
ig\xi_0 I_3 &  2g^2\xi_0^2/\omega_0 I_3 & 0\\
ig/\xi_0 & 0 & 0 
\end{pmatrix}   
\end{equation}
with $I_3$ the $3\times3$ unit matrix, representing the independent dipole polarizations.

\begin{equation}
\alpha_V=\begin{pmatrix}
 |\mathbf k+\mathbf G|& 0 \\
0 & -|\mathbf k+\mathbf G|,
\end{pmatrix}
\end{equation}
is the a $2N_G\times2N_G$ matrix with the energies of the  scalar and longitudinal photons for each $\mathbf G$ (polarized along $\mathbf k+\mathbf G$). The dipoles also interact with the transverse photons with wavevector $\mathbf k+\mathbf G$. In the Coulomb gauge these are the free photons and their interaction with matter occurs through light-matter coupling. For our purpose, only longitudinal and scalar photons are considered into the self-energy correction of the dipole excitations. Finally, the coupling matrices are written as
\begin{equation}
\kappa_1=\kappa_2=ig\begin{pmatrix} \xi_\mathbf{G}  (\hat e_1 \cdot \hat n_{\mathbf k+\mathbf G}) & (\hat e_1 \cdot \hat n_{\mathbf k+\mathbf G})/\xi_\mathbf{G} \\\xi_\mathbf{G}  (\hat e_2 \cdot \hat n_{\mathbf k+\mathbf G}) &  (\hat e_2 \cdot \hat n_{\mathbf k+\mathbf G})/\xi_\mathbf{G} & \\
 \xi_\mathbf{G} (\hat e_3 \cdot \hat n_{\mathbf k+\mathbf G}) & (\hat e_3 \cdot \hat n_{\mathbf k+\mathbf G})/\xi_\mathbf{G} & \dots\\
0 & 0  \\ 
 0& 0\\ 
 0& 0\\ 
 0& 0
\end{pmatrix},
\end{equation}
After some algebra, we find the self-energy $\Delta_{\mathbf k}=\kappa\mathcal G_V \bar\kappa$ to be 
\begin{equation}
\Delta_{\mathbf k}=\begin{pmatrix}
\bar\Delta & \bar\Delta \\
-\bar\Delta & -\bar\Delta \end{pmatrix} 
\end{equation}
with
\begin{equation}
\bar\Delta=
\begin{pmatrix}
\delta_1 & 0 & 0 & 0 & 0 & 0 & 0\\
0 & \delta_2 & 0 & 0 & 0 & 0 & 0\\
0 & 0 & \delta_3 & 0 & 0 & 0 & 0\\
0 & 0 & 0 & 0 & 0 & 0 & 0\\
0 & 0 & 0 & 0 & 0 & 0 & 0
\end{pmatrix}
\end{equation}
and
\begin{equation}
\delta{j}=\frac{2g^2}{\omega_0} \sum_{\mathbf G\neq0}   \frac{\omega_0^2-|\mathbf k+\mathbf G|^2}{|\mathbf k +\mathbf G|^2-\omega^2}\left(\hat e_j \cdot \hat n_{\mathbf k-\mathbf G}\right)^2.\label{eq:deltaj}
\end{equation}
In the particular case of a cubic lattice, we get that $\delta_1=\delta_2=\delta_3=\delta_{\mathbf k}(\omega)$, where $\delta_{\mathbf k}(\omega)$ is given in Eq.~(\ref{eq:deltak}). 

Substitution of these corrections onto the $D^{corr}_{\mathbf k}=D_\mathbf k-\Delta_{k}$ leads to a Dynamical matrix which can be separated into three independent parts: two identical parts corresponding to the transverse excitations, and one corresponding to the longitunal excitations. These independent dynamical matrices are shown in the main text in Eqs.~(\ref{eq:DLT}), (\ref{eq:alpgammaT}) and (\ref{eq:alpgammaL}).

\section{Diagonalization method \label{app:diag}}

In order to find the transformation $\tilde T_{\mathbf k}$ which preserves the metric $\Upsilon$ for the longitudinal part of the Hamiltonian, we first diagonalize the dynamical matrix numerically by a similarity (conjugation) transformation. The obtained right eigenvector $V_{\mathbf k}$ is first renormalized by $\tilde V_{\mathbf k}=V_{\mathbf k}(V_{\mathbf k}^\dag \Sigma V_{\mathbf k})^{-1/2}$. With this transformation the dynamical matrix can be brought into a block diagonal form $\tilde D_{\mathbf k}=\Sigma\tilde V_{\mathbf k}^\dag\Sigma D_{\mathbf k} \tilde V_{\mathbf k}$ where
\begin{equation}
\tilde D=\begin{pmatrix}
\omega_L &  0 & 0 & 0 \\
0 & \mathcal D &  0 & 0\\
0 & 0 & -\mathcal D^\dag & 0 \\
0 & 0 &  0 & -\omega_L
\end{pmatrix}.  
\end{equation}
For $k<k_\mathrm{EP}$ we have $\mathcal D= \begin{pmatrix} \mathrm{Im}(\omega_\pm) & \mathrm{Re}(\omega_\pm) e^{i\theta} \\ \mathrm{Re}(\omega_\pm) e^{-i\theta} & -\mathrm{Im}(\omega_\pm)  \end{pmatrix}$; $\omega_\pm$ and $\theta$ are the magnitude and phase of the imaginary eigenvalues of the numerically diagonalized dynamical matrix
\begin{equation}
\Omega_k=\mathrm{diag}(\omega_L,~\omega_+,~\omega_-,-\omega_-,-\omega_+,-\omega_L).    
\end{equation}
 
The numerical diagonalization of the dynamical matrix leads to a non-diagonal metric $\tilde\Upsilon =\tilde V_{\mathbf k}^\dag \Sigma \tilde V_{\mathbf k}$ for the commutation relation  
\begin{equation}
\tilde\Upsilon=\begin{pmatrix}
1 &  0 & 0 & 0 \\
0 & \mathcal I &  0 & 0\\
0 & 0 & -\mathcal I^\dag & 0 \\
0 & 0 &  0 & -1
\end{pmatrix}    
\end{equation}
with $\mathcal I=\begin{pmatrix} 0 & e^{i\beta} \\  e^{-i\beta} & 0  \end{pmatrix}$. 

To obtain the generalized BV transformation, we proceed by numerically diagonalizing $\mathcal I$ separately and using the resulting eigenvectors to obtain a transformation that follows the original commutation relations ($\tilde T_{\mathbf k}^\dag \Sigma \tilde T_{\mathbf k}=\Upsilon$). For $k<k_\mathrm{EP}$, the obtained generalized BV transformation fails to diagonalize the longitudinal part of the dynamical matrix, yielding a block diagonal matrix with the block $\tilde{\mathcal D}=\begin{pmatrix} -\mathrm{Re}(\omega_\pm) & \mathrm{Im}(\omega_\pm) \\ \mathrm{Im}(\omega_\pm)  & \mathrm{Re}(\omega_\pm)  \end{pmatrix}$. Clearly, one can choose to either diagonalize the dynamical matrix, leading to linearly independent solutions which do not follow the original commutation relations or diagonalize the metric, restoring the original commutation relations, but leading to linearly dependent excitations. In this work we chose the later in order to allow for a well defined Fock state on which the subsitidary conditions could be imposed. 

For $k>k_\mathrm{EP}$, $\mathcal D$ and $\mathcal I$ are already diagonal and this procedure does not affect the resulting dynamical matrices and commutation relations.  

\section{Subsidiary condition \label{app:subs}}

Here we show how the obtained BV transformation can be used to impose the subsidiary condition in Eq.(\ref{eq_subs1}). We start by defining the contribution of each hybrid state $\psi_\lambda(\mathbf k)$ to the subsidiary condition 
\begin{equation}
Z_\lambda(\mathbf k)=[\left(kA_L-[A_0,\mathcal H]\right),\psi_{\lambda,\mathbf k}^\dag]\label{eq_subs1},    
\end{equation}
and writing the transformation matrix \begin{equation}
\tilde T=\begin{pmatrix}\tilde T^{11} & \tilde T^{12} \\ \tilde T^{21} & \tilde T^{22} \end{pmatrix}  , 
\end{equation}
such that the bare states $\phi_{s,\mathbf k}$ can be expressed in terms of the hybridized states
\begin{equation}
\left\{\begin{matrix} \phi_{s,\mathbf k}=\sum_\lambda \tilde T^{11}_{s,\lambda}\psi_{\lambda,\mathbf k}+\tilde T^{12}_{s,\lambda}\psi_{\lambda,-\mathbf k}^\dag\\
\phi_{s,-\mathbf k}^\dag=\sum_\lambda \tilde T^{21}_{s,\lambda}\psi_{\lambda,\mathbf k}+\tilde T^{22}_{s,\lambda}\psi_{\lambda,-\mathbf k}^\dag
\end{matrix} \right..\end{equation}
Here $s=LD,LP,SP$, for the longitudinal dipole excitation ($LD$), the longitudinal photon ($LP$) and the scalar photon ($SP$), while $\lambda=L,+,-$ for the three hybrid solutions. 

We note that $[\phi_{s,\mathbf k},\mathcal H]= \sum_{s'} D^{11}_{ss'}(\mathbf k)\phi_{s,\mathbf k}+D^{12}_{ss'}(\mathbf k)\phi_{s',-\mathbf k}^\dag$, and $[\phi_{s,-\mathbf k}^\dag,\mathcal H]= \sum_{s'} D^{21}_{ss'}(\mathbf k)\phi_{s',\mathbf k}+D^{22}_{ss'}(\mathbf k)\phi_{s',-\mathbf k}^\dag$, where $D$ here denotes the Dynamical matrix for the $G=0$ longitudinal part of the Hamiltonian  (already including the self-energy correction). Substituting these into Eq.(\ref{eq_subs1}) we have
\begin{equation}
Z_\lambda(\mathbf k)=k\left(\tilde T^{11}_{2,\lambda}(\mathbf k)+\tilde T^{21}_{2,\lambda}(\mathbf k)\right) - \sum_{s'}\left(D^{21}_{ss'}(\mathbf k)\tilde T^{11}_{s',\lambda}(\mathbf k)+D^{22}_{ss'}(\mathbf k)\tilde T^{21}_{s',\lambda}(\mathbf k)\right).
\end{equation}

\bibliographystyle{unsrt}
\bibliography{LT_paper,LT_paper_mendeley}

\end{document}